\documentclass[aps,prl,twocolumn,amsmath,amssymb]{revtex4} 
\usepackage[english]{babel}
\usepackage{latexsym}
\usepackage{graphics}
\usepackage{subfigure}
\usepackage{epsfig}
\begin{document}

\title{Three-dimensional quasi-Tonks gas in a harmonic trap}
\author{P. Pedri$^{1,2}$, and L. Santos$^{1}$}  
\address{(1) Institut f\"ur Theoretische Physik, Universit\"at Hannover, D-30167 Hannover,Germany}
\address{(2) Dipartimento di Fisica, Universit\`a di Trento and BEC-INFM, I-38050 Povo, Italy}
\begin{abstract}  

%
%
We analyze the macroscopic dynamics of a Bose gas in a harmonic trap with a superimposed 
two-dimensional optical lattice, assuming a weak coupling between different lattice sites.
We consider the situation in which the local chemical potential at each lattice site 
can be considered as that provided by the Lieb-Liniger solution. 
Due to the weak coupling between sites and the form of the chemical potential, 
the three-dimensional ground-state density profile and the excitation spectrum 
acquire remarkable properties different from both 1D and 3D gases.
We call this system a quasi-Tonks gas. We discuss the range 
of applicability of this regime, as well as realistic experimental situations where it can be observed.
\end{abstract}  
\pacs{03.75.Fi,05.30.Jp} 
\maketitle


Recent developments in low-dimensional trapping \cite{1D}, 
loading of optical lattices \cite{Latt,Greiner}, and external modification of the 
interparticle interactions \cite{Feshbach}, open fascinating perspectives 
towards the achievement of strongly-interacting systems in cold atomic gases. 
In this sense, a remarkable example is provided by the 
recent observation of the superfluid to Mott-Insulator (MI) transition 
in cold bosonic gases in optical lattices \cite{Greiner}. Other strongly-interacting 
systems in cold atomic gases have been recently considered theoretically, as the case of large 
scattering length \cite{biga}, rapidly rotating Bose gases \cite{belen}, and 
one-dimensional systems, both bosonic \cite{Olshanii2,Chiara,Letter,Gangardt,Giorgini} 
and fermionic \cite{Recati}.

One-dimensional systems differ significantly when compared to 3D 
ones. Most remarkably, in 1D gases the interactions become relatively more important 
the more dilute the gas is. In particular, for very low densities or a large value of the 
$s$-wave scattering length $a$, the system enters the so-called Tonks-Girardeau (TG) regime, 
in which the bosons acquire fermionic properties \cite{Girardeau60,Wright}. For 
finite interactions, the problem of delta-interacting 1D homogeneous Bose gases 
was exactly solved by Lieb and Liniger (LL) by means of Bethe Ansatz \cite{LiebLiniger}. 
A combination of the LL solution, the local density approximation, and hydrodynamic equations 
have been recently employed in the analysis of different properties of trapped 1D Bose gases, as 
the ground-state density \cite{Olshanii2}, the excitation spectrum \cite{Chiara}, 
the non self-similar expansion in a 1D guide \cite{Letter}, and 
the correlation and coherence properties \cite{Gangardt}. The latter has also been recently analyzed 
by means of quantum Monte Carlo methods \cite{Giorgini}.

The accomplishment of strongly-interacting 1D gases requires 
tight transversal trapping, low atom numbers, and possibly the increase 
of $a$ via Feshbach resonances \cite{Olshanii2,Olshanii1,Petrov1D}. 
In this sense, 2D optical lattices are specially favorable, 
since the on-site transversal confinement can be very strong. 
For sufficiently intense lattices, the tunneling 
rate could be made small enough to ensure that each lattice site 
behaves as an independent 1D experiment. 
This paper analyzes the situation in which the tunneling 
is non negligible, but sufficiently weak to assume the   
1D character of the local chemical potential.
In this regime, which we call the quasi-Tonks regime, we predict that the combination 
of tunneling and 1D local chemical potential results in 
important modifications of the 3D ground-state density and the 
excitations.


In the following we consider a gas of $N$ bosons in a 
cylindrically-symmetric harmonic trap, with 
axial (radial) frequency $\omega_z$ ($\omega_\perp$). We  
assume a superimposed periodic potential provided by an optical lattice of the form 
$V_{\rm l}(x)+V_{\rm l}(y)=V_0(\sin^2qx+\sin^2qy)$, where
$q=2\pi/\lambda$ is related to the laser wavelength $\lambda$, 
and $V_0$ is the lattice amplitude. 
The lattice has a periodicity 
$d=\pi/q=\lambda/2$. From a macroscopic point of view it is convenient to introduce 
the effective mass, $m^*$, to describe the effects of the lattice 
on the dynamics. The value of $m^*$ can be obtained 
from $\partial^2E_0/\partial k^2=\hbar^2/m^*$, where 
$E_0(k)$ is the dispersion law corresponding to the lowest energy band. In tight-binding regime, 
$m^*$ can be related with the tunneling rate $J$, as 
$m/m^*=\pi^2J/E_R$, where $E_R=\hbar^2q^2/2m$ is the recoil energy, 
$J=-\int w_i(x)(-\hbar^2\nabla_{x}^2/2m +V_{\rm l}(x)) w_{i+1}(x)dx$, and   
$\{w_i\}$ are the Wannier functions for the lowest band.


We first consider the case of a single isolated lattice site, which can be well 
approximated as a dilute gas of $N_t$ bosons 
confined in a very elongated harmonic trap with radial and axial frequencies 
$\omega_\rho$ and $\omega_z$ ($\omega_\rho\gg\omega_z$). 
The transversal confinement is strong enough to fulfill that the 
interaction energy per particle is smaller than the 
zero-point energy $\hbar\omega_\rho$ of the transversal trap. In this way, the 
transversal dynamics is effectively ``frozen'' and the 
system can be considered 1D. 
The Hamiltonian describing the physics of the delta-interacting 1D gas is
\begin{equation}
\hat{H}_{\rm 1D}=\hat{H}_{\rm 1D}^{0} + 
\sum_{j=i}^{N_t}\frac{m\omega_{z}^{2} z_{i}^{2}}{2}
\end{equation}
where
\begin{equation}
\hat{H}_{\rm 1D}^{0}=
-\frac{\hbar^{2}}{2m}\sum_{j=1}^{N_t}\frac{\partial^{2}}{\partial z_{j}^{2}}
+ g_{\rm 1D}\sum_{i=1}^{N_t-1}\sum_{j=i+1}^{N_t}\delta \left(z_{i}-z_{j}\right)
\end{equation}
is the Hamiltonian in absence of the harmonic trap,  
and $g_{\rm 1D}=-2\hbar^{2}/ma_{\rm 1D}$. 
The scattering problem under one-dimensional constraints was analyzed in detail 
by Olshanii \cite{Olshanii1}, and it is characterized by 
the 1D scattering length 
$a_{\rm 1D}=(-a^{2}_{\rho}/2 a)[1-{\mathcal C}(a/a_{\rho})]$, 
with $a$ the 3D scattering length, $a_\rho=\sqrt{2\hbar/m\omega_\rho}$   
the oscillator length in the radial direction, and ${\mathcal C}=1.4603\dots$.
For the thermodynamic limit, a 1D gas at zero temperature with linear density $n$, 
is characterized by an energy per particle, which can be obtained from the 
corresponding LL integral equations \cite{LiebLiniger}. 
Assuming that the density variates sufficiently slowly, 
at each point $z$ the gas can be considered in local 
equilibrium, and the local energy per particle is provided by the LL equations 
for the density $n(z)$. We call this approach Local Lieb-Liniger (LLL) model.


Although, strictly speaking, the LLL approach is only valid for 1D systems, 
for sufficiently low $J$, the virtual processes involving transitions into nearest sites, 
should just result in a small correction to the 1D scattering of the order of $J/\mu$, 
where $\mu$ is the chemical potential. Therefore, if $J/\mu\ll 1$, the actual local chemical potential 
can be well approximated by that obtained from the LLL approach. 
If the tunneling becomes very small, even for large values of $N_t$ 
the system could enter into the MI regime. However, 
this regime typically demands an extremely small value of $J/\mu$. Summarizing, in 
what we called the quasi-Tonks regime, 
$1/N_t\lesssim J/\mu\ll 1$, the LLL chemical potential can be employed, and 
at the same time the tunneling cannot be neglected.


Assuming a sufficiently slow variation of the gas density, the dynamics 
can be well described by means of the corresponding macroscopic hydrodynamic equations 
\cite{meret}: 
\begin{eqnarray}
\frac{\partial n}{\partial t}+\nabla\cdot(n{\bf v})=0 \label{conteq}\\
m^*\frac{\partial v_x}{\partial t}+\frac{\partial}{\partial x}(K+V+U)=0\\
m^*\frac{\partial v_y}{\partial t}+\frac{\partial}{\partial y}(K+V+U)=0\\
m\frac{\partial v_z}{\partial t}+\frac{\partial}{\partial z}(K+V+U)=0
\end{eqnarray}
where
\begin{eqnarray}
K&=&\frac{m^*}{2}v_x^2+\frac{m^*}{2}v_y^2+\frac{m}{2}v_z^2\\
V&=&\frac{m}{2}\omega^2_\perp(x^2+y^2)+\frac{m}{2}\omega^2_zz^2\\
U&=&\mu(n)=\mu_{1D}(d^2n)
\label{U}
\end{eqnarray}
where $n(x,y,z)$ is the macroscopic 3D density and ${\bf v}=(v_x,v_y,v_z)$ is the velocity field. Note
that all the information about the lattice is contained in 
$m^*$, and in the local chemical potential $\mu(n)$.
The use of the hydrodynamic equations is justified by the
superfluidity of the quasi-1D tubes \cite{Sonin} and by the avoidance of 
the MI phase. 
In the equation (\ref{U}) the local chemical potential of the 3D system
is related to the 1D chemical potential, $\mu_{1D}$, provided by the LLL approach. 
In order to obtain $\mu_{1D}$ the value of $a_{1D}$ must be obtained for each site. 
The latter demands the knowledge of the effective oscillator length $a_\rho$ associated 
with the on-site radial confinement. Since the latter can be very large, 
a Gaussian Ansatz can be assumed for the on-site radial wave function. In this way, we can  
calculate $a_\rho$ by minimizing the radial energy
\begin{eqnarray}
\frac{E(a_\rho)}{E_R}=\frac{1}{q^2a_\rho^2}+\frac{V_0}{E_R}\left(1-e^{-q^2a_\rho^2}\right). 
\label{radialE} 
\end{eqnarray}
In Eq.~(\ref{radialE}) we have neglected the interaction energy, 
which for $V_0/E_R>3$ does not provide any significant contribution. 
Once known the value of $a_{1D}$, we evaluate $\mu_{1D}(n)$ by numerically 
solving the corresponding LL integral equations.


\begin{figure}[ht!]
\begin{center}\
\psfig{file=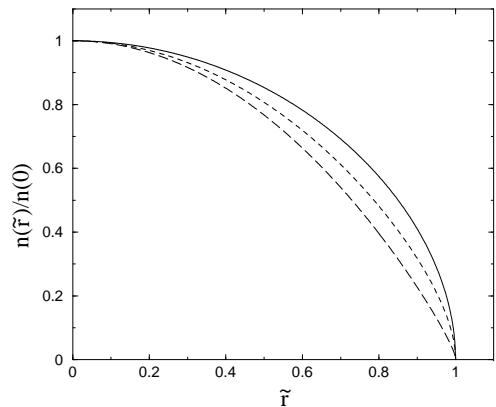,width=6.5cm}\\[0.1cm]
\end{center} 
\caption{Density profile as a function of the dimensionless radius 
$\tilde{r}^2=\sum_i (r_i/R_{i})^2$ (see text). The solid, doted and dashed lines
correspond to $\log\,A=-3$, $0$, $3$, 
respectively. The resulting fit with a power law dependence 
$(1-\tilde r^2)^s$ provides $s=0.54$, $0.72$, $0.91$, respectively, being an indicator of the transition 
from the MF regime to the strongly-interacting one.}
\label{density}  
\end{figure}

By imposing the equilibrium conditions $\partial n/\partial t=v_i=0$, we 
obtain the equation of state
\begin{equation}
\mu[n_0(x,y,z)]=\mu_T - V(x,y,z) ,
\label{eqst}
\end{equation}
where $n_0$ is the equilibrium density, and $\mu_T$ is the chemical 
potential of the system. 
Inverting (\ref{eqst}), one obtains the expression for $n_0$.
Imposing the normalization to the number of particles $N$ we obtain:
\begin{eqnarray}
A=\frac{|a_{1D}|^4d^2N}{a_x^2a_\perp^4}=4\pi\xi^3\int_0^1 t^2\tilde{
\mu}_{1D}^{-1}[\xi^2(1-t^2)]dt, 
\label{A}
\end{eqnarray}
where $\xi=2\mu m|a_{1D}|^2/\hbar^2$. From 
Eq.~(\ref{A}) we observe that, similarly to the strict 1D case 
\cite{Olshanii2,Chiara,Gangardt}, the ground state of the system is completely 
characterized by a single parameter $A$. For $A\gg 1$ the mean-field (MF) regime 
is retrieved, whereas the TG is found for $A\ll 1$. In Eq.~(\ref{A}), 
$\tilde{\mu}_{1D}(x)=2m|a_{1D}|^2\mu_{1D}/\hbar^2$, is a  
function of the dimensionless density $\eta=|a_{1D}|d^2n$. Fig.~\ref{density} 
shows the density as a function of the dimensionless radius $\tilde{r}$, where 
$\tilde{r}^2=\sum_i (r_i/R_{i})^2$, with $R_i$ the Thomas-Fermi radii. 
In order to compare the different cases, we re-scale the density with respect 
to the central one. Nevertheless, we stress that the values of $R_i$ and the 
central density depend on the parameters, and in particular on the 
interaction regime. 

\begin{figure}[ht!]
\begin{center}\
\psfig{file=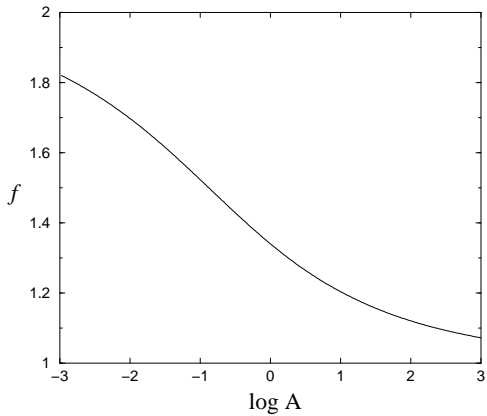,width=6.5cm}\\[0.1cm]
\end{center} 
\caption{Parameter $f$ as a function of $\log\,A$ (see text).}
\label{fpara}  
\end{figure}

From the previous discussion, it becomes clear that the 
1D properties of the local chemical potential
are transferred to the 3D stationary shape of the cloud. 
In particular, for small values of $A$, the transversal density profile is not an 
inverted parabola, as expected for the case of a BEC in a harmonic trap. We stress once more, 
that this remarkable effect is just possible in the presence of the 2D lattice, 
which guarantees, under the above discussed conditions, a LLL chemical potential. 

%
In the following we calculate the frequencies for the lowest excitations 
of a quasi-Tonks gas. In particular, we evaluate 
the breathing and quadrupole modes, by using the 
Ansatz $n=n_0\left(r_i/b_i\right)/\prod_jb_j$, $v_i=(\dot{b}_i/b_i)r_i$, which 
constitutes an exact solution of the continuity equation 
(\ref{conteq}). Multiplying the Euler equations by $nx_i$ and 
integrating, one obtains
\begin{eqnarray}
\ddot{b_x}+\tilde{\omega}_\perp^2b_x+
\frac{\tilde{\omega}_\perp^2}{b_x}F(\prod_{j=x,y,z} b_j)=0\\
\ddot{b_y}+\tilde{\omega}_\perp^2b_y+
\frac{\tilde{\omega}_\perp^2}{b_y}F(\prod_{j=x,y,z} b_j)=0\\
\ddot{b_z}+\omega_z^2b_z+\frac{\omega_z^2}{b_z}F(\prod_{j=x,y,z} b_j)=0,
\end{eqnarray}
where we have employed the effective frequencies
$\tilde{\omega}^2_\perp=\omega^2_\perp m/m^*$
and defined the function 
\begin{equation}
F(\rho)=\frac{1}{m\omega_i^2N\langle x_i^2\rangle_0}
\int n_0x_i\frac{\partial}{\partial x_i}\mu(\rho^{-1}n_0)d^3r.
\end{equation}
Using the previously calculated equilibrium solution,  
one obtains that $F(\rho)$ is independent of the 
particular choice of the coordinate $i$, and that $F(1)=-1$. 
Note that the function $F$ only depends on $A$, which, as previously 
discussed, completely characterizes the static properties of the system. 
Linearizing around the equilibrium solution $b_i=1$, one 
obtains the frequencies 
of the breathing mode ($\omega_B$), the $m=0$ quadrupole mode ($\omega_{Q1}$) 
and the $m=2$ quadrupole one ($\omega_{Q2}$): 
\begin{widetext}
\begin{eqnarray}
\omega^2_B&=& \frac{1}{2}\left(\tilde{\omega}_\perp^2(2+2f)+\omega_z^2(2+f)+
\sqrt{ [\tilde{\omega}_\perp^2(2+2f)+\omega_z^2(2+f)]^2-8 \tilde{\omega}^2
_\perp\omega_z^2(2+3f)
}
\right)
\label{wB}
\\
\omega^2_{Q1}&=& 
\frac{1}{2}\left(\tilde{\omega}_\perp^2(2+2f)+\omega_z^2(2+f)-
\sqrt{ [\tilde{\omega}_\perp^2(2+2f)+\omega_z^2(2+f)]^2-8 \tilde{\omega}^2
_\perp\omega_z^2(2+3f)
}
\right)
\label{wQ1}
\\
\omega^2_{Q2}&=& 2\tilde{\omega}_\perp^2
\label{wQ2}
\end{eqnarray}
\end{widetext}
In the expression for the frequencies we employ the parameter $f=F'(1)$, which 
is an universal function of the parameter $A$. We show this dependence
in Fig. \ref{fpara}. The parameter $f$ ranges from $2$, at the 
TG limit, to $1$, for the MF regime. In this two limiting cases the function
$F$ can be obtained analytically, being $F=\rho^{-2}$ (TG) and 
$F=\rho^{-1}$ (MF).


\begin{figure}[ht] 
\begin{center}\
\psfig{file=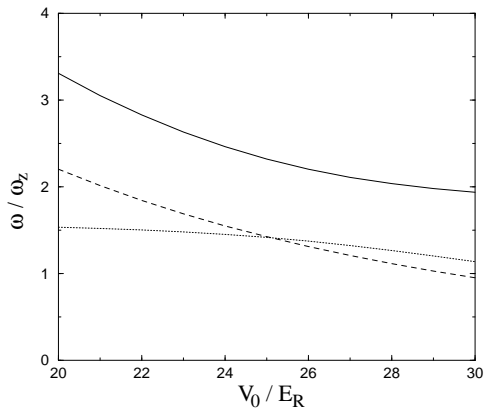,width=6.5cm}\\[0.1cm]
\end{center} 
\caption{Frequencies of the breathing mode (solid), quadrupole $1$ (dotted) and 
quadrupole $2$ (dashed) as a function of $V_0/E_R$, for the particular case of  
 $^{87}$Rb atoms, $N=2\times 10^{5}$, 
$\omega_z=2\pi\times 4$Hz, $\omega_\perp=2\pi\times 40$Hz, and $d=0.5\mu$m. 
}
\label{exper}  
\end{figure}
%


The quasi Tonks regime is not only achievable for realistic conditions, 
but, actually, it is expected to be the case for typical 
parameters in ongoing experiments, as those of Fig.~3, which shows the dependence of 
the frequencies (\ref{wB}-\ref{wQ2}) on the lattice amplitude $V_0/E_R$. 
We stress that the system enters the MI regime 
for large values of $V_0/E_R$ (around $40$ for the case of Fig.~3). If this is the case, as discussed above, 
our macroscopic hydrodynamic approach fails, and the system becomes a set of independent 1D gases. 
On the other hand, for decreasing values of $V_0/E_R$ the system abandons the quasi-Tonks regime 
(which for the case of Fig.~\ref{exper} occurs at $V_0/E_R\simeq 15$), and the condition  
$J/\mu\simeq 1$ is reached. This constitutes an additional 
intermediate cross-dimensional regime, in which the gas is 3D, but the 
local chemical potential is not that expected for a 3D Bose gas. For this regime, contrary 
to the case of the quasi-Tonks gas, the local chemical potential cannot be approximated 
by the LLL chemical potential. As a consequence of that, the challenging problem of describing
the $J/\mu\simeq 1$ regime demands a completely different approach. In this sense, especially 
interesting  
could be to employ the analogies to the problem of coupled Luttinger liquids \cite{Luttinger}.


The existence of the quasi Tonks regime can be easily revealed in experiments, by 
either observing the size and the form of the stationary density profile, 
or monitoring the collective excitations. 
For increasing values of $V_0/E_R$, the frequency of the breathing mode approaches that of 
the lowest compressional mode in 1D gases \cite{Chiara}, whereas the quadrupolar frequencies 
tend to zero. For intermediate values of the lattice potential, the lowest excitations are 
significantly different than in 1D, for the same value of the dimensionless density $\eta$.
In particular, more than one excitation mode is expected. We stress that the modes 
(\ref{wB}-\ref{wQ2}) also differ significantly from the expected results for a BEC in a lattice 
\cite{meret}. The latter are recovered from our formalism for large values of $A$ (MF regime).
We must point, however, that the local correlation properties are preserved 
in the quasi Tonks regime. In this sense, if the parameter $A$ becomes sufficiently small 
the lifetime of the gas can be very significantly enlarged, due to the 
reduction of the two- and three-body losses \cite{Gangardt}.


In this Letter, we have considered what we call the quasi-Tonks regime, 
in which a gas confined in a 2D optical lattice, can present significant tunneling, 
and at the same time maintain the local chemical potential obtained for each site using the LLL model.   
We have discussed that this regime should be the one encountered under current 
typical experimental 
conditions. In this cross-dimensional regime, a 3D cloud acquires 1D properties. In particular, we 
have shown that both the ground-state properties and the excitation spectrum are 
significantly different than the expected results for a harmonically confined BEC.

We thank M. Lewenstein, P. \"Ohberg, L. P. Pitaevskii, 
G. V. Shlyapnikov, S. Stringari, and the experimental Quantum Optics 
group of the University of Hannover for
stimulating discussions.
We acknowledge support from Deutsche Forschungsgemeinschaft SFB 407 and SPP1116,  
the RTN Cold Quantum Gases, IST Program EQUIP, ESF PESC BEC2000+, EPSRC, 
and the Ministero dell'Istruzione, 
dell'Universit\`a e della Ricerca (MIUR).
The authors wish to thank the Alexander von Humboldt Foundation, 
the Federal Ministry of Education and 
Research and the ZIP Programme of the German Government.

\end{document}